\documentstyle[aas2pp4]{article}
\def\alwaysmath#1{\ifmmode{#1}\else{$#1$}\fi}

\def\msun{\alwaysmath{\,{M}_{\odot}}}

\def\feh{{\rm [Fe/H]}}
\def\ofe{{\rm [O/Fe]}}
\def\alphafe{{\rm [\alwaysmath{\alpha}/Fe]}}
\def\metals{{\rm [\alwaysmath{M}/H]}}

\newcommand\hst{{\it HST}}

\newcommand\etal{et~al.}
\newcommand\msto{MSTO}
\newcommand\fig[1]{Figure~\protect\ref{#1}}
\slugcomment{ApJ, in press}

\lefthead{Rood et al.}
\righthead{Luminosity Function of M3}

\begin{document}

 \title{The Luminosity Function of M3\footnote{Based in part on
observations with the NASA/ESA {\it Hubble Space Telescope}, obtained at
the Space Telescope Science Institute, which is operated by AURA, Inc.,
under NASA contract NAS5-26555}}

\author{R.T. Rood\altaffilmark{2},
E. Carretta\altaffilmark{3},  
B. Paltrinieri\altaffilmark{4},  
F. R. Ferraro\altaffilmark{5,6}
}

\author{F. Fusi Pecci\altaffilmark{6,7},
B. Dorman\altaffilmark{8},
}

\author{A. Chieffi\altaffilmark{9},
O. Straniero\altaffilmark{10},
R. Buonanno\altaffilmark{11}
}

\altaffiltext{2}{Astronomy Dept, University of Virginia,
	P.O.Box 3818, Charlottesville, VA 22903-0818; rtr@virginia.edu}
\altaffiltext{3}{Osservatorio Astronomico di Padova, vicolo dell'Osservatorio 5,
 35122 Padova, ITALY; carretta@astrpd.pd.astro.it}
\altaffiltext{4}{Istituto di Astronomia-Universit\'a "La Sapienza",
via G.M. Lancisi 29, I-00161 Roma, Italy; barbara@coma.mporzio.astro.it}
\altaffiltext{5}{European Southern Observatory, Karl Schwarzschild Strasse 2,
D-85748 Garching bei M\"unchen, Germany}
\altaffiltext{6}{Osservatorio Astronomico di Bologna, via Ranzani 1, 40126
Bologna, ITALY; ferraro,flavio@astbo3.bo.astro.it}
\altaffiltext{7}{Stazione Astronomica di Cagliari, 09012 Capoterra, ITALY}
\altaffiltext{8}{Laboratory for Astronomy \& Solar Physics, 
Code 681, NASA/GSFC,	Greenbelt MD 20771; dorman@veris.gsfc.nasa.gov}
\altaffiltext{9}{Istituto di Astrofisica Spaziale-CNR, 00044 Frascati,
	ITALY; alessandro@altachiara.ias.rm.cnr.it}
\altaffiltext{10}{Osservatorio Astronomico di Collurania, 64100
	Teramo, ITALY; straniero@astrte.te.astro.it}
\altaffiltext{11}{Osservatorio Astronomico di Roma, 00040 Monte Porzio, ITALY; buonanno@coma.mporzio.astro.it}

\begin{abstract}

We present a high precision, large sample luminosity function (LF) for the
Galactic globular cluster M3. With a combination of ground based
and {\it Hubble Space Telescope} data we cover the entire radial extent of
the cluster. The observed LF is well fit by canonical
standard stellar models from the red giant branch (RGB) tip to below the main
sequence turnoff point. Specifically, neither the RGB LF-bump nor
subgiant branch LF indicate any breakdown in the standard models. On
the main sequence we find evidence for a flat initial mass function
and for mass segregation due to the dynamical evolution of the cluster.

\end{abstract}


\section{Introduction}

One of the classical tests of stellar evolutionary calculations is
through the comparison of theoretical and observed stellar luminosity
functions (LFs) of globular clusters (e.g., \cite{rfp88}). The LF for
the stars below the main sequence turn-off (\msto) can be related to
the stellar initial mass function and be used as a probe of stellar
dynamics within a cluster. The LF of stars after the \msto\ depends
primarily on the rate of evolution and provides a direct and
straight-forward test of evolutionary calculations. Basically the
post-\msto\ LF measures the development of the hydrogen burning shell
and its progression through the star. Features in the observed LF can
be related to interior structure. For example, as first noted by Iben
(1968), when the H-burning shell advances through the composition
discontinuity left by the maximum penetration of convective envelope,
the luminosity briefly decreases. This leads to the so-called LF
bump. More subtle observables are also present: the slope of the LF
below and above the LF bump are different because the H-burning shell
is in the first case passing through a region of varying H abundance
and later through a region of constant H (\cite{rc85},
\cite{fpbump90}). A breakdown in {\em canonical} stellar evolution
theory (\cite{rfp88}) can affect the LF in the region of the subgiant
branch (SGB).\footnote{The {\em canonical} assumptions include the
neglect of phenomena like rotation, non-convective matter transport,
etc. in stellar model calculations.  Some of these are certainly
present in real stars, but they are very difficult to include in
calculations. This leads to two approaches to dealing with
non-canonical phenomena: one is to try to model the phenomenon
ordinarily in a simplistic way requiring the introduction of free
parameters. The other is to make the most stringent possible
comparison between observations and canonical models---the approach we
adopt here.} Examples include extra energy transport by Weakly
Interacting Massive Particles (WIMPs) (\cite{fs93}) or the
gravitational settling of helium (\cite{pv91}, \cite{cast97},
Straniero \etal\ 1997), both of which affect the cluster age
calibration.

To explore short evolutionary phases or to search for the small
changes in the LF produced by reasonable modifications to the
canonical models requires large samples.  Indeed, to explore phenomena
like those cited above, such large samples are required that
photometry must extend to the crowded inner parts of clusters. To be
useful in constructing a LF, photometric samples must be complete and
accurate---a requirement which has come into conflict with the
necessity of working near cluster cores. Adequate samples to identify
the LF-bump were first obtained for 47 Tuc by King, DaCosta \&
Demarque (1985) and for 11 other clusters by Fusi Pecci \etal\ (1990)
(see Cassisi \& Salaris 1997 for a review). At lower luminosities LF
studies have produced some puzzling results. Stetson (1991) in a LF
formed from combined data from the low metallicity clusters M68, M92,
and NGC~6397 found an excess of stars on the SGB just above the \msto.
Bolte (1994) found a similar result for M30 which has been 
confirmed by Bergbusch (1996), Guhathakurta \etal\ (1998) and Sandquist \etal\
(1999). This excess was especially intriguing because that was the
``signature'' expected if WIMP energy transport was important
(\cite{fs93}). However \cite{sbsh96} in the largest sample LF obtained
to date did not find a similar result for the somewhat more metal rich
cluster M5.

Clearly larger and better LFs are desirable, and these are now
possible with the {\it Hubble Space Telescope} (\hst\/) which has the
capability to obtain photometry near cluster centers. While very
large \hst\ samples of fainter stars can be obtained, the small field
of view limits the number of evolved stars which can be observed. However,
the combination of ground based photometry of the outer cluster and \hst\
data can yield a large sample LF which spans the range of cluster stellar
luminosities. In this paper we present such an LF
obtained for the cluster M3 (NGC~5272). This is the largest, most precise,
and most complete LF yet obtained for a globular cluster.

\section{HST-WFPC2 Data Sample}

\subsection{Observations and Data Analysis}

The observations were obtained April 25,  1995 as part of the \hst\
program GO5496 (P.I. F. Fusi Pecci), the first of our
programs devoted to testing the theoretical
models using luminosity functions of a set of Galactic Globular
Clusters with different metallicity.

A set  of exposures was secured through UV (F255W and F336W) and
optical filters (F555W and F814W). Preliminary results on UV data has been
already presented in Ferraro \etal\ (1997a) and  \cite{lag98};
here we will report on results
obtained using only the $V$ and $I$ exposures.
Total exposure times were 400 sec in $V$ (F555W) and 560 sec in
$I$ (F814W); each exposure is listed in Table 1 by Ferraro \etal\ (1997a). 

The initial data analysis (including bias, dark and flat-field
corrections) was made through the standard $HST$ pipeline.  A master
unsharp mask image for each filter was made using MIDAS to register
all the long-exposure images, yielding a median frame with cosmic rays
statistically eliminated.  The search of the individual star
components was made on the $V$ median frame, following the usual
procedure with ROMAFOT (Buonanno \etal\ 1983).  Then we performed the
PSF fitting on each individual frame separately using a Moffat
function plus a numerical map of the residuals to better account for
the contribution of the stellar PSF wings.  The average instrumental
magnitudes for each star were transformed to the standard Johnson $V$
and $I$ system using the \cite{holt1995} recipes.  The magnitudes thus
obtained were eventually shifted to match the independent zero-point
calibration adopted in Ferraro et al. (1997b, hereafter M3CCD).  The
$V$,~$V-I$ CMD for the HST global sample is shown in \fig{hstcmd}.

Special care has been devoted to ensure that the effects of blends
(and crowding) do not significantly alter the derived LF.  The effects
of blending can be quite problematic in CMDs of dense regions of
globular clusters. On the main sequence, blended stars can potentially
be mistaken for binary stars, while on the subgiant branch they be
shifted erroneously to brighter bins. Blends can mimic
incompleteness by shifting the two
stars off the main loci such that they are dropped from the
LF. 
To minimize these effects
suspected blended objects with multiple stellar components lying off
the main branches of the CMD were individually examined by eye,
exploiting the fully interactive capability of the package ROMAFOT.  In
this way we were able to check and, if necessary,
eliminate spurious objects. The degree to which blends contaminate the
final sample can be estimated from the artificial stars tests
described below. Some  of the artificial stars which were not
recovered were lost as blends. Thus the degree of incompleteness at a
given level serves as an upper limit to the blend-fraction.

\subsection{Completeness}

A crucial component in the calculation of an accurate LF is the
determination of corrections for incompleteness.  In our data crowding
is the primary source of incompleteness.  Moreover, because of the
large gradient in the spatial density of stars present in our frame,
the completeness at any level of magnitude must be determined as a
function of the position with respect to the cluster center.

In order to quantify the efficiency in detecting stars as function of
the magnitude and the degree of crowding we carried out extensive
artificial star tests in the $V$ filter frames,
following the procedure described below.

\begin{enumerate}

\item Each chip was split into two annular
regions centered on the cluster center:

\begin{description}
\item{INNER:} $r<50\arcsec$ including the entire PC1 field and the innermost 
part
of the three WFC fields;

\item{OUTER:} $r>50\arcsec$ covering the outer part of the three WFCs.

\end{description}

M3 is a cluster of moderate central density. With a core radius $r_c =
30\arcsec$ (Djorgovski 1993) one would expect the stellar density to be
approximately constant across the PC as is apparent from
\fig{WFPC2fld}. However, the stellar density has dropped appreciably at the
outer parts of the WFs. We have adopted 
50\arcsec\ as a boundary between regions of higher and lower projected
stellar density.  A map of the WFPC2 field of view is shown
\fig{WFPC2fld} with the circle (at $r \sim 50\arcsec$) delineating the two
regions.

 \item A set of artificial stars was generated using the {\it
Tiny~Tim} package (Krist 1994), and these were added to the original
frames in random positions.  In
order to avoid a spurious crowding enhancement, the size of the added
sample (about 300 total artificial images or $\sim 80$ for each chip)
was $\leq 10$\% of the total number of stars originally found
in each chip. Then the star searching and fitting procedures were
repeated on the artifically enriched frames.

 \item An artificial
star was considered to be ``recovered'' on the basis of the following
criteria:

\begin{itemize}
 \item the differences in position, $\Delta x,~\Delta y$, between the
input artificial star and the recovered one are $\Delta
x,~\Delta y<1$ pixel;

\item the difference in magnitude, $\Delta mag$, is $|\Delta mag| <0.3$.

\end{itemize}
The fraction of artificial stars missed during any step of the
procedure is a good estimate of the actual number of stars missed at
each magnitude level during the reduction procedure. 

\item The completeness factor ($\Phi$) 
was finally computed as the numbers
of stars recovered ($N_{rec}$) with respect to the number 
of stars simulated ($N_{sim}$) during each trial.
The final value of $\Phi$ has been obtained as an average of 3--6 independent
trials for each magnitude bin (0.2 mag width) in each of the two regions
(INNER and OUTER).

\end{enumerate}

The result of these tests is summarized in \fig{complt}a,b. The
completeness at the \msto\ level ($V\sim 19$) is $>90\%$ in both the
INNER and OUTER regions, thus the star counts up to that level have
been corrected by less than $10\%$.  The corrections for completeness
were applied separately in each part (inner or outer) of each field
(PC, WFCs) so that the total corrected sample properly accounts for
the differing degrees of crowding.

\subsection{The Mean Ridge Line of the RGB, SGB, and Upper MS}

If there were no photometric errors all of the RGB, SGB, and Upper MS
stars would lie along a line in the CMD. We approximate this line with
a  mean ridge line (MRL) which  
has been determined
by an iterative procedure:

\begin{enumerate}

\item A first rough selection of stars belonging to the RGB, SGB, MS was 
performed by eye, excluding the horizontal branch (HB), asymptotic
giant branch (AGB), and blue straggler stars (BSS);

\item This preliminary sample was divided into 0.5 mag width bins;

\item For each bin the median $V-I$ color was computed; 

\item All stars lying more than 5$\sigma_{\scriptscriptstyle V-I}$ from
the median value was rejected. 

\end{enumerate}

Points (3) and (4) of the procedure were iterated until the solution was
stable. The resulting MRL points are listed in Table 1. Only
about 2\% of the sample was rejected in determining the MRL. 

\begin{deluxetable}{lcclccclc}
\small
\tablewidth{\textwidth}
\label{ridgeline}
\tablecaption{Mean ridge line for the HST sample.}
\tablehead{
 \colhead{$V$}           & \colhead{$V-I$} &\colhead{}&
\colhead{$V$} & \colhead{$V-I$} &\colhead{} & \colhead{} &
\colhead{$V$} & \colhead{$V-I$}
}
\startdata
{\bf RGB} &  &   &  &   &    &    & {\bf HB}  & \nl
   12.44 &    1.589 &   &   18.16 &    0.803 &  &   &  15.57 &    0.775\nl
   12.68 &    1.473 &   &   18.26 &    0.787 &  &   &  15.62 &    0.731\nl
   12.90 &    1.403 &   &   18.33 &    0.764 &  &   &  15.63 &    0.684\nl
   13.10 &    1.347 &   &   18.37 &    0.746 &  &   &  15.64 &    0.631\nl
   13.30 &    1.294 &   &   18.42 &    0.724 &  &   &  15.64 &    0.585\nl
   13.55 &    1.241  &   &  18.47 &    0.696 &  &   &  15.62 &    0.272\nl
   13.79 &    1.193 &   &   18.52 &    0.668 &  &   &  15.63 &    0.219\nl
   14.05 &    1.150 &   &   18.57 &    0.647 &  &   &  15.64 &    0.184\nl
   14.28 &    1.115  &   &  18.62 &    0.629 &  &   &  15.66 &    0.146\nl
   14.56 &    1.077 &   &   18.70 &    0.616 &  &   &  15.71 &    0.105\nl
   14.82 &    1.045  &  &   18.78 &    0.608 &  &   &  15.76 &    0.079\nl
   15.09 &    1.017  &   &  18.86 &    0.600 &  &   &  15.83 &    0.043\nl
   15.36 &    0.989 &   &   18.98 &    0.593 &  &   &  15.91 &    0.008\nl
   15.64 &    0.964 &  &    19.10 &    0.592 &  &   &  15.99 &   -0.021\nl
   15.89 &    0.939 &   &   19.18 &    0.593 &  &   &  16.10 &   -0.050\nl
   16.21 &    0.911 &   &   19.26 &    0.595 &  &   &  16.21 &   -0.069\nl
   16.52 &    0.890  &   &  19.38 &    0.597 &  &   &  16.37 &   -0.080\nl
   16.70 &    0.879  &   &  19.78 &    0.624 &  &   &  16.54 &   -0.088\nl
   16.94 &    0.867  &   &  20.10 &    0.654 &  &   &  16.65 &   -0.091\nl
   17.20 &    0.856  &   &  20.30 &    0.672 &  &   &  &  \nl
   17.36 &    0.849  &   &  20.50 &    0.693 &  &    & &  \nl
   17.63 &    0.836 &   &   20.98 &    0.756 &  &    & &  \nl
   18.05 &    0.815  &   &  21.34 &    0.805 &  &   &  &  \nl
\enddata
\end{deluxetable}

 Luminosity functions were constructed for each chip
by dividing each individual CMD in 0.2 mag bins, and simply
counting all the stars in each bin within $\pm
5\sigma_{\scriptscriptstyle V-I}$ 
of the MRL.

\fig{hstlf} shows a zoom of luminosity function in the Turn-Off region 
for the total HST sample, before and after correction for completeness.
Errors have been computed accordingly to the following 
relation (see also Bolte 1989):

\begin{equation}
\sigma_{\rm bin}=[{N^{1/2} \over \Phi} + {N \cdot \sigma_\Phi
   \over \Phi^2}]
\end{equation}

\noindent where $N$ is the number of stars  observed in
each bin, $\Phi$ is the completeness factor, and $\sigma_\Phi$ is the
associated error which was determined from the rms of the repeated
completeness trials.

\section{Matching the Ground-Based Observations: the Global LF}

$HST$ data presented here complement the large data set we have
obtained for M3 during the last decades.  The photographic sample was
originally presented by Buonanno {et al.} (1994) and the CCD ground
based data have been discussed by Ferraro \etal\ (1993, 1997b).  The
entire photographic sample extends over an area between 2\arcmin\ and
7\arcmin\ from the cluster center. Here we consider a subsample
(hereafter referred to as PHOTO) of stars lying in the annulus between
$3\farcm 5 < r < 6\arcmin$ from the cluster center (region AI in
Buonanno {et al.} 1994).  In this annulus all stars down $V\sim22$
have been measured.  The CCD data cover a field of view $7\arcmin
\times 7\arcmin$ roughly centered on the cluster center; stars in the
innermost central regions ($r<20\arcsec$) have been not measured.  In
this sample the deeper exposures ($V$ and $I$) only reached to $V \sim
20$.  In order to ensure an high level (formally $100\%$) of
completeness we include only stars with $V<17$ lying in the annulus
100\arcsec--210\arcsec\ from the cluster center. We will refer to this
subsample as CCD. The area covered by the CCD subsample and the the
others fields in shown Figure 1 of Ferraro \etal\ (1997b).

In the PHOTO and CCD subsets (in analogy to the HST sample) we applied the
procedure described in Section 2.2, selecting {\it bona fide} RGB-SGB-MS
members taking all stars within $\pm 5\sigma$ in color from the MRL in the
$V,~B-V$ and $V,~V-I$ CMDs respectively.  The CMDs of all stars selected
in the three samples (PHOTO, CCD, HST respectively) are shown in
\fig{3samplecmds}. For each of these samples we independently derive the
LF.  Star counts have been corrected using the respective completeness
curve: the PHOTO sample has been corrected using Table 7 by Buonanno et al.
(1994), the CCD sample (for $V<17$) is virtually complete, thus no
correction has been applied, and the HST sample has been
corrected using the completeness curves plotted in \fig{complt}a,b. In
constructing the global LF we considered only the magnitude range where
the completeness was greater than 50\%.  Thus the global LF is restricted
to $V \la 21$ by the most incomplete region in the HST sample---the inner
region of the WF4 and WF2 cameras (\fig{complt}a,b).

Since the CCD sample
spanned a different magnitude range than the others, we determined a
scaling factor to produce a proper match.  We followed the
procedure explained in Bolte (1994): let $n_{i,k}$ be the number of
stars, after the proper completeness correction, in  bin $i$ of the
sub-sample $k$, where $k=1,2,3$  indicates the PHOTO, the global
$HST$, and the CCD sub-samples, respectively.  Thus, for $k=1,2$ the
resulting LF is simply given by the sum

\begin{displaymath}
N_i^{'} = \sum_{k=1}^{2} n_{i,k}
\end{displaymath}
in the range $12 < V < 21$.
However, in order to include the CCD sub-sample ($k = 3$) we derived a scaling 
factor $s$ from the range in magnitude ($12 < V < 17.0$) which is
common to all sub-samples:

\begin{displaymath} 
s = {\sum_{k=1}^{2} n_{i,k} \sum_{i} n_{i,k} \over 
\sum_{k=1}^{3} n_{i,k} \sum_{i} n_{i,k}}
\end{displaymath} 

\noindent so that the number of stars for each bin in the combined LF is

\begin{displaymath} 
N_i = s \times (N_i^{'} + n_{i,k=3}).
\end{displaymath} 

Table 2 lists the final LF.  Column 1 lists the magnitude bins (which
are 0.2 mag. in width); column 2 gives $N_i$, the total scaled number of
objects per 0.2 mag; column 3 contains the errors in the $N_i$.  Even
though it extends only 2 mag below the turnoff, this
LF, including more than 50000 stars, is the most populated LF ever
published for a galactic globular cluster.

\begin{deluxetable}{lccclcc}
\small
\tablewidth{\textwidth}
\label{globallf}
\tablecaption{Luminosity function of the global sample (PHOTO+CCD+HST).}
\tablehead{
 \colhead{$V$}           & \colhead{$Log(N_i)$}      &
\colhead{$\sigma N_i$} &  \colhead{} & \colhead{$V$}&  
\colhead{$Log(N_i)$}&\colhead{$\sigma N_i$}
}
\startdata
    12.70 &     0.70 &     0.45 &  & 17.50 &     2.05 &     1.10\nl
    12.90 &     0.30 &     0.26 &  & 17.70 &     2.22 &     1.20\nl
    13.10 &     0.78 &     0.45 &  & 17.90 &     2.26 &     1.23\nl
    13.30 &     0.85 &     0.41 &  & 18.10 &     2.46 &     1.35\nl
    13.50 &     0.95 &     0.52 &  & 18.30 &     2.72 &     1.52\nl
    13.70 &     0.85 &     0.50 &  & 18.50 &     2.97 &     1.68\nl
    13.90 &     0.90 &     0.50 &  & 18.70 &     3.06 &     1.75\nl
    14.10 &     0.90 &     0.47 &  & 18.90 &     3.18 &     1.83\nl
    14.30 &     1.11 &     0.61 &  & 19.10 &     3.25 &     1.88\nl
    14.50 &     1.08 &     0.55 &  & 19.30 &     3.32 &     1.94\nl
    14.70 &     1.32 &     0.67 &  & 19.50 &     3.37 &     1.98\nl
    14.90 &     1.15 &     0.58 &  & 19.70 &     3.42 &     2.01\nl
    15.10 &     1.34 &     0.77 &  & 19.90 &     3.45 &     2.04\nl
    15.30 &     1.38 &     0.72 &  & 20.10 &     3.49 &     2.08\nl
    15.50 &     1.76 &     0.92 &  & 20.30 &     3.53 &     2.12\nl
    15.70 &     1.60 &     0.87 &  & 20.50 &     3.56 &     2.14\nl
    15.90 &     1.51 &     0.75 &  & 20.70 &     3.56 &     2.16\nl
    16.10 &     1.65 &     0.83 &  & 20.90 &     3.59 &     2.19\nl
    16.30 &     1.70 &     0.93 &  & 21.10 &     3.58 &     2.21\nl
    16.50 &     1.79 &     0.96 &  & 21.30 &     3.58 &     2.24\nl
    16.70 &     1.85 &     1.00 &  & 21.50 &     3.55 &     2.27\nl
    16.90 &     1.95 &     1.06 &  & 21.70 &     3.57 &     2.35\nl
    17.10 &     2.04 &     1.09 &  & 21.90 &     3.66 &     2.48\nl
    17.30 &     2.06 &     1.11 &  & &  & \
\enddata
\end{deluxetable}

It is useful to compare our LF to that for M5 obtained by Sandquist
\etal\ (1996, hereafter SBSH), the best previous globular cluster
LF. Doing this, it is appropriate to compare sample size in specific
parts of the CMD. Our ``bright'' sample which corresponds to the upper
4.5 mag of the RGB (or $V \le 17$) contains 944 stars. For such bright
stars SBSH were able to work near the cluster center, and their sample
in the comparable RGB interval was about 1000 stars. For fainter stars
ground based photometry becomes increasingly affected by problems of
photometric blends (see SBSH). Blends can be especially insidious when
they ``move'' a star from one part of the LF to another, because this
is just the sort of thing one would expect in a breakdown of canonical
stellar modeling. To avoid blends only the sparsely populated outer
cluster can be used, and it is especially difficult to get adequate
samples of the SGB region. Here our blend-free HST sample plays a
crucial role.  We have 6085 stars brighter than \msto; SBSH have 3300.
In a 0.1 mag bin at turnoff we have 820 stars compared to SBSH's 280.

\section{Results}

\subsection{Preliminaries}

There are certain issues which must be addressed in comparing LFs
either with other observational LFs or with theoretical LFs. The first
is how the samples are normalized. Beyond this there is some interplay
between assumed distance modulus, age, and chemical composition.

\subsubsection{Normalization}

One must determine a ``normalization'' factor to adjust for
sample size. In recent LF studies (e.g.
\cite{stetson91,bolte94}, Sandquist \etal\ 1996) both the lower RGB and the main
sequence region have been used for normalization. 
We adopt a different approach
normalizing to the total number of stars brighter than the \msto\ at $V=19.10$.
Operationally this has the advantage that the turnoff is well defined
and because of the large number of stars involved the normalization is
not much affected by Poisson statistics.
The small imprecision in determining the location of the
turnoff ($\la 0.1\,$mag) produces only a very small error ($\la 1$\%)
in the normalization.

\subsubsection{ Chemical composition}

For M3 Kraft \etal\ (1992), found $\feh = -1.47$, based on detailed
fine abundance analysis of high resolution spectra of red giant
stars. However, their study is not fully self-consistent since while
stellar analysis was performed using model atmospheres from the grid
of Bell et al. (1976, hereafter BEGN), the reference solar iron
abundance was extracted from the empirical Holweger and M\"uller
(1974, hereafter HM) solar model.  This inconsistency was resolved by
Carretta \& Gratton (1997) who use the Kurucz (1993) grid of model
atmospheres for both solar and stellar analysis, giving $\feh =
-1.34$. The offset with the study of Kraft \etal\ (from which the
observational material was taken) is simply due to the fact that the
HM model is $\sim 150$\,K warmer than the BEGN models in the line
formation region, explaining the 0.13 dex difference in the derived
[Fe/H].

Oxygen in red giants can by affected by interior nuclear
processing. Indeed, M3 is one of the several clusters with an observed
anti-correlation between O and Na along the RGB (see for instance
Kraft \etal\ 1992). This anti-correlation arises because of
non-standard mixing on the upper RGB. Therefore, the best estimate for
the $\ofe$ ratio is that obtained considering only unmixed stars,
i.e. giants with $\rm [Na/Fe] \approx 0.0$ dex. Using Kraft \etal\
(1992) we estimate that $\ofe = 0.19 \pm 0.04$ dex based on 5 stars.

One typically assumes that the other $\alpha$-elements (e.g. Kraft
\etal\ 1993) have the same enhancement as O and uses two metallicity
parameters, \feh\ and \alphafe, in computing models. For many problems
one can get by with just one metallicity parameter using the short cut
of Salaris \etal\ (1993). This makes use of the {\it global}
metallicity (\metals) based on the mass fraction of all elements
heavier than helium as compared to solar. They found that scaled solar
models with $\feh = \metals$ mimicked models based on observed \feh\
with $\alpha$-enhancement. 
For M3 $\metals= -1.2$. So while ideally
we would compare M3 data to models with $\feh = -1.34$ and $\ofe =
0.2$, a comparison to scaled solar models with $\feh = \metals =
-1.2$ should be an adequate surrogate. Given typical errors in
metallicity determinations it is appropriate to explore a region $\pm
0.3\,$dex from these values.

\subsubsection{Distance modulus and age} 

In principle, fitting cluster main sequences to subdwarfs with precise
{\it Hipparcos} parallaxes should lead to relatively high precision
distance moduli. However, M3 was not included in the first round of
clusters with distances determined in this way
(\cite{reid97,gratton97}).  The main reason was that the fiducial
cluster main sequence ridge line (from Ferraro et al. 1997b; CCD97) in
the $V,~B-V$ plane did not reach deep enough with the photometric
accuracy required to apply sub-dwarf fitting to determine distance
moduli. However, using an expanded sample of local subdwarfs from the
newly available Hipparcos catalogue (Carretta et al. 1999a) it
is possible to give an estimate for the distance modulus of M3
in the framework of the longer distance scale defined by Gratton
\etal\ (1997) and Reid (1997). With $E(B-V)=0.02\pm 0.01$ and
$\feh=-1.30$ (to be consistent with reddening and metallicity scales
used in Gratton \etal\ 1997), the true distance modulus for M3 is
$(m-M)_0 = 15.13$ (Gratton 1998, private communication, adopting
the Ferraro et al 1997b photometry). 
A cautionary flag to the accuracy of sub-dwarf fitting is set by the
fact that different groups using this method do not get the same
result (Pont \etal\ 1997; Reid 1997; Gratton
\etal\ 1997).  Small details are important (Carretta \etal\ 1999a), and
systematic effects could leave residual errors of 0.2--0.3 mag in
$(m-M)$.

There are other independent estimates of the distance to M3. Cudworth
(1979) gives a kinematic distance for M3 of $(m-M)_0 = 14.91$. From the
RR~Lyrae Sandage and Cacciari (1990) give values of $(m-M)_0$ ranging from
14.81 to 15.00 depending on which metallicity/absolute magnitude relation
is adopted for the RR~Lyrae. Using the latest metallicity/absolute
magnitude determined from Baade-Wesselink distances to field RR~Lyrae
(Fernley \etal\ 1998) and our adopted metallicity ($\metals = -1.2$), we
get $(m-M)_0 = 14.86$ (assuming $E(B-V)=0.01$). In a recent analysis of
the RR~Lyrae distance scale, De Santis (1996) gives $(m-M)_0 = 15.03$.
Marconi \etal\ (1998) find that $(m-M)_0 = 14.94$ gives a good fit to the
LF of the lower main sequence.  Ferraro \etal\ (1999) have recently
obtained $(m-M)_0 = 15.05$ within the framework of a homogeneous
re-analysis of the evolved sequences of the CMD in a sample of about 60
GGCs.  They derived the distance modulus from the comparison of the
observed level of the ZAHB ($V_{ZAHB}$) and the theoretical models
computed by SCL97. Their use of synthetic HBs to determine the ``observed''
$V_{ZAHB}$ is a significant improvement over earlier applications of the
same technique.

 Given the present state of uncertainty we consider it reasonable to
consider $(m-M)_V = 14.8$--15.2. The question of age is intertwined
with that of distance. The longer distance scales imply smaller
ages. For example, with the ``sub-dwarf distance'' given above and
using stellar models from Straniero \etal\ 1997 (private
communication), the corresponding absolute age for M3 is about 12 Gyr.
An decrease in $(m-M)$ of 0.07\,mag leads to an increase in age of
1\,Gyr. Hence ages of 11--15\,Gyr for M3 are not beyond question.

\subsection{The Red Giant Branch}

The differential  LF for our global sample is shown
in \fig{lftot}. The dominant determinant of RGB-LF is simply energy
conservation via the so-call energy consumption theorem
(\cite{fuelcon}). Basically the number of stars in some interval along
the RGB, $N_i$, depends inversely on the luminosity at that point,
$N_i \propto 1/L$. So $\log N_i \propto M_{bol}$, and neglecting
bolometric corrections $\log N_i$ should drop linearly with magnitude
moving up the RGB just as seen.

Theoretical evolutionary models predict the existence of a ``feature''
in the RGB called the {\it LF-bump}.  This LF-bump marks the
evolutionary stage at which the H-burning shell passes through the
composition discontinuity left by the maximum penetration of the
convective envelope (\cite{ibenbump,rfp88}).  As emphasized first
by Rood \& Crocker (1985), the practical detection of such a feature
requires  very large RGB samples.  Because of
this, our global sample gives us the possibility to obtain one of the
most precise determinations of the LF-bump.

Models also predict that in the integrated LF the slope changes on either side
of LF-bump. The slope change results from the fact that the
H-burning shell is moving through a region of increasing hydrogen
abundance below the bump and constant hydrogen abundance above the
bump. This change in slope can more reliably locate the bump
than observations of the bump itself (Rood \& Crocker 1985; Fusi Pecci
\etal\ 1990). 

In \fig{lftot} the LF-bump is very well defined in both the
differential and integrated LF, and is located at $V=15.45\pm0.05$ (in
good agreement with the previous determinations (Buonanno \etal\ 1994,
Ferraro \etal\ 1997b). Also note in \fig{lftot} the nomenclature we use
to refer to the different branches. Our nomenclature is that of the
theorist in which the branches are defined in terms of the supposed
interior structure. There is often confusion, in particular, over the
term sub-giant branch. To the theorist, and in this paper, SGB refers to
the transition region between the MSTO and the lower RGB where the
ridge line in the CMD is more horizontal $V=18.2$--18.6. Structurally this is the
branch of thick H-shell burning where the degenerate core
develops. For historical reasons observers often refer to stars along the
lower-RGB as sub-giants. 

  We compare our results to two new sets of theoretical evolutionary
tracks/isochrones/LFs which use the new OPAL opacities and equation of
state (e.g., Rogers \& Iglesias 1992). This should offer significant
improvement over earlier models particularly for opacity dependent LF
features like the LF-bump. The first set (the no-diffusion models
described in Straniero \etal\ 1997, hereafter SCL97) is available for
scaled-solar abundances and may be used for any given value of
$\alpha$ enhancement with the Salaris \etal\ (1993) prescription. The
second (Bergbush \& VandenBerg 1999, herafter V97) has $\alphafe =
0.3$. The V97 models also include improved color-temperature
relations. On the RGB the V97 models utilize a non-Langrangian
numerical technique which is not optimal for investigations of the RGB
bump. This technique might be expected to smear and perhaps slightly
lower the LF-bump.

In \fig{cs-van} the global differential LF is compared to SCL97 and
V97 models for $(m-M)_V=15.03$, $\rm age = 12\,Gyr$. For SCL97 models
we took $\feh =
\metals = -1.2$. The V97 models have $\feh=-1.41$ and $\alphafe =
0.3$ which is equivalent to $\metals = -1.2$. The integrated LF
compared to SCL97 is shown in \fig{lfint}. The fits are really quite
good with the region of poorest fit at $V \sim 18$ just above the SGB.
That the theoretical and observed LFs agree so well for stars brighter
than $V \sim 18$ shows that the hydrogen profile in the region $M(r)
\sim 0.2$--0.4\msun\ is given fairly accurately by canonical models.

The values of $(m-M)$, age, and metallicity which lead to the best LF
fit have the virtues of also agreeing well with the longer Hipparcos
distance scale with their correspondingly shorter ages and with the
abundances obtained by Carretta \& Gratton (1998). Furthermore, these
parameters fit in well with those obtained from a new re-analysis of
the location of the LF-bump in a sample of more than 40 GGCs by
\cite{bump99}. They find that the previous discrepancy in the bump
location between the
observations and models is completely removed using new models and
considering the newer global metallicity scale.

The good fit to the LF-bump comes at the expense of some mismatch at
the SGB/lower-RGB transition. To explore how varying the parameters
affect the fit we must quantify the quality of the fits.  Formally we
can perform a Chi-squared test computing

\begin{displaymath}
\chi^2 = \sum_{i=1}^{N} \left({{\log(N_{\rm observed})_i - 
\log(N_{\rm model})_i} \over {\log(N_{\rm observed})_i }}\right)^2
\end{displaymath}

\noindent For the best fit to the entire data set ($13<V<20.5$), shown
in \fig{cs-van}, there is a slightly better agreement with the SCL97
models ($\chi_{\rm SCL97}^2 = 0.11$ and $\chi_{\rm V97}^2 = 0.13$).

One could conceivably prefer parameters which do not give the best
overall fit. The factors which determine the luminosity of the
``theoretical'' LF-bump have been discussed in Fusi Pecci \etal\
(1990), Chieffi and Gratton (1986) and more recently and extensively
in \cite{cs97}. The most important factors are the opacity in the
1--$2 \times 10^6\,$K range and the degree that convective elements
undershoot the neutral buoyancy point of the convective
envelope. Elsewhere the post-TO LF depends mostly on the rate of fuel
consumption (but see \S \ref{sec:sbg} below), and one might expect it
to be a considerably more robust prediction of the models than the
location of the LF-bump. Given this one might tolerate some residual
error in the LF-bump location to achieve a better fit elsewhere and,
thus, exclude the bump region when measuring the quality of the fit. If we
do so for the fits shown in \fig{cs-van} we find $\chi_{\rm
SCL97\_xbump}^2=0.01$ and $\chi_{\rm V97\_xbump}^2=0.02$.

In \fig{scl} we explore the effect of varying the distance modulus,
age and chemical composition. Since the quality of the fit is
essentially the same for SCL97 and V97, we show results only for the
SCL97 models. We have found it impossible to fix parameters so that
the model fits the SGB region and the LF-bump equally well. The best
fit in the SGB region results from a higher age of 14\,Gyr.  However
the good fit in the SGB region brings with it a worse fit for the
LF-bump. The fit to the LF-bump can be restored at an age of 14\,Gyr
by adopting a higher metallicity and $(m-M)$ smaller than we consider
plausible (\fig{scl}b). The upper end of the allowable $(m-M)$ range
gives good fit to the LF-bump (\fig{scl}c) at the expense of a
slightly worse fit to the SGB region and a small age (10\,Gyr).  The
Chi-square values both including and excluding the LF-bump region are
given in Table 3.

\begin{deluxetable}{lccc}
\tablewidth{\textwidth}
\label{chi2}
\tablecaption{Chi-square results between observations and SCL models}
\tablehead{
 \colhead{}           & \colhead{$13<V<17$}      &
\colhead{$17<V<20.5$} & \colhead{$13<V<20.5$}
}
\startdata
$\rm [M/H]=-1.2 $& & &\nl
$(m-M)_V=15.03 $& & &\nl
$\rm age=12\,Gyr$& 0.101 & 0.014 & 0.114\nl
 & & & \nl
$\rm [M/H]=-1.2 $& & &\nl
$(m-M)_V=14.80 $& & &\nl
$\rm age=14\,Gyr$& 0.194 & 0.008 & 0.201\nl
 & & & \nl
$\rm  [M/H]=-0.82 $& & &\nl
$(m-M)_V=14.65 $& & &\nl
$\rm age=14\,Gyr$& 0.103 & 0.026 & 0.130\nl
 & & & \nl
$\rm [M/H]=-1.3 $& & &\nl
$(m-M)_V=15.20 $& & &\nl
$\rm age=10\,Gyr$& 0.112 & 0.018 & 0.131\nl
 & & & \nl

\enddata
\end{deluxetable}

From the $\chi$-values listed in Table 3 we conclude that considering both
the two main observables of the LF (the LF-bump and the SGB-rise) we can
find a good agreement (with both SCL97 and V97 models) when we take $\rm
age=12$, $(m-M)_V = 15.03$ and $[M/H]\sim -1.2$ (see \fig{cs-van}).  On
the other hand, if we consider that some residual error still might be
present in the theoretical prediction of the LF-bump location,
we  get the best agreement between theory and
observations if $\rm age=14$, $(m-M)_V = 14.80$ and
$[M/H]\sim -1.2$ (see \fig{scl}a). 

A full discussion of the age should also bring in the isochrone fits.
These are complicated both by the color calibration and the
color-temperature relations. Probably because of this our isochrone fits
using the ages and distances determined from the LF give no further
indication of whether the age of M3 is closer to 12 or 14 Gyr.

\subsection{The Lower Red Giant Branch and  Sub-Giant Branch\label{sec:sbg}}

The lower-RGB \& SGB LF can be especially interesting. This is the
region in which there has been some indication of a breakdown in the
canonical models for very low metallicity clusters (\cite{stetson91},
\cite{bolte94}) and recently confirmed for M30
(\cite{bergbuschm30,guham30,sandqm30}).  It is the region affected by
non-canonical assumptions like WIMP energy transport (\cite{fs93},
\cite{dwl97}), and helium settling (\cite{pv91,scl97}), both of which
could reduce cluster age estimates. Rapid stellar rotation
(VandenBerg, Larsen, \& de Propris 1998) and \alphafe\ could also
affect the SGB region of the LF. We illustrate all of these in
\fig{sgblf} (see also Figs. 2 \& 3 of Vandenberg \etal\ 1998).  The
predicted changes are all small, and age differences can mimic these
changes.

As shown in Figs.~\ref{cs-van} \& \ref{lfint}, the the theoretical and
observed LFs agree well in the SGB region and lower-RGB. Indeed the
fits are pretty good under a wide range of assumptions (\fig{scl}). We
should note that for the excess of lower-RGB stars in M30 cited above,
the observed LF lies above the lower-RGB at essential every bin. For
M3 our ``bad'' fits correspond to only one or two mismatched
bins. The fits can be essentially perfect (\fig{scl}a) if we accept
some error in the bump location. Basically, there is no indication for
a breakdown in the canonical models. SBSH found a similar result for
M5.

On the other hand the observational consequences of plausible
non-canonical effects are small so even our good fit may not place
very tight constraints on their magnitude. We do not attempt to place
such constraints now because we feel that a traditional LF analysis
may not be the appropriate way to analyze the SGB. By projecting the
star count information onto the ``magnitude'' axis one loses the
information contained in the color distribution of the stars. In a
future paper we will report on our efforts to study the distribution
``along'' the SGB (see for example, \cite{bv97}), and provide
quantitative estimates as to the degree that non-canonical assumptions
can be ruled out.

\subsection{Below the Main Sequence Turnoff}

Below the \msto\ the LF primarily probes the stellar initial mass
function (IMF) and cluster dynamics. Having been acquired primarily to
study post-\msto\ evolution, the data we describe in this paper
reaches only $\sim 2\,$mag below the turnoff which corresponds to a
mass significantly above the lower mass cutoff for main sequence
stars.  Deeper LFs (based on almost 25,000 stars) exploring this
region are  presented in Marconi \etal\ (1998) and Carretta \etal\ (1999c).

It is widely recognized that stars in dense systems (like GGCs) are
subject to energy exchange through stellar encounters and rapidly
evolve toward a state of equipartition of energy (so that stars of
lower mass will have higher velocities).  As a direct consequence
low mass stars should have larger average distance from
the center and their distribution will be less centrally concentrated
than that of higher mass stars. This mass segregation effect is expected to
produce an observable effect on the LF if a sufficiently large radial
region of the cluster is sampled.

Such mass segregation has already been found in GGCs
surveyed by HST: NGC~6397 (King \etal\ 1995), 47~Tuc (Paresce \etal\
1995 ), M15 (De Marchi \& Paresce 1996), NGC\,6752 (Ferraro \etal\
1997c). Piotto \etal\ (1997) give an extended discussion of this
topic. Mass segregation has been found in M30 using ground based data
(Bolte 1989). 

Even with our limited MS coverage, we can see evidence for mass
segregation in M3.  In \fig{lf3radii} we show a comparison of the LFs
in three separate radial regions: $r < 20\arcsec$; $50\arcsec < r <
100\arcsec$; $210\arcsec < r < 360\arcsec$.
As above the LFs have been normalized 
to the total number of stars brighter than the MSTO.
 The inner regions show a
significant drop of star counts over the magnitude range $V\sim
20-21.5$.  We interpret the observed depletion as due to the mass
segregation effect. The depletion is obvious even in the small
interval of masses sampled.  The effect shown in \fig{lf3radii} cannot
due to residual incompleteness of the samples.  In order to completely
remove the effect we would have to assume a level of completeness
which is far smaller than the estimated value.  For example in the
innermost sample $r<20''$ (which is almost completely contained in the
PC) the measured completeness must be almost a factor 2 in error
(dropping from 0.8 down to 0.45) in order to remove the effect.

Is the observed decrease in the star counts consistent with the
theoretical expectation?  In order to derive the relative distribution
of stars of different mass at different distances from the cluster
center we have used the formula given by King \etal\ (1995) following
the procedure as explained in
DeMarchi and Paresce (1996). The faint end of the LF should be depleted
in the core region with respect to the outer regions by the factor $
f = (\rho_0/\rho_{\rm ext})^{m_2/m_1 -1 } $ where $\rho_0$ and
$\rho_{\rm ext}$ are the stellar densities in the core and in the
external region respectively, and $m_2$ and $m_1$ are the stellar masses
at the faint extreme and  TO level of the LF,
respectively. We use the tables of SCL97 and assume $(m-M)_V=15.0$ to relate
observed magnitudes to mass---the \msto\ at $V=19.1$ corresponds to
$M_V\sim4$ and a mass of $\sim 0.84$\msun, $V=20.5$ corresponds to
$M_V=5.5$ and a mass of $\sim 0.73$\msun, and $V=21.9$
corresponds $M_V=6.95$ and a mass of $\sim 0.61$\msun, respectively.

The spatial densities can be approximated by the Eqn. 5 in King \etal\ 
(1995). In the case of M3 the external region lies at $r_{\rm ext}\sim
290\arcsec$, and taking $r_c \sim 30\arcsec$ (Djorgovski 1993), we
estimate that the ratio $(\rho_0/\rho_{\rm ext})$ is $\sim 900$.
Adopting the mass values obtained above, the density of low-mass stars
should be a factor $900^{0.13}\sim 2.4$ less than that in the central.
This value is compatible with the observed depletion which is
$975/548\sim 1.8$.

We can also explore the stellar initial mass function (IMF) using our
data. The IMF is normally approximated by a power-law mass spectrum of the
form $\phi(M)dM=M^{-(1+x)}dM$ where $\phi(M)dM$ is the number of stars in
the mass range $M$ to $M+dM$.  The observed slope of the IMF measured in
terms of $x$ should vary with radius in a way consistent with a underlying
global slope.  To verify this we use the dynamical models of Pryor, Smith,
\& McClure (1986).  We compute $x$ using Eq.~3 of Bolte (1989).  The two
magnitude intervals used are $18.55<V<20.55$ and $20.55<V<21.95$.
\fig{pryormod0} shows the observed $x$ plotted as a function of radius.
We see that $x({\rm observed})$ varies $r$ in a way consistent with the
$x({\rm global}) \sim 0$. This is close to the value we inferred from our
total sample.

\section{Summary}

We have obtained a $V$-band luminousity function for the globular
cluster M3 using a combination of ground based photographic and CCD
observations and \hst\ observations of the cluster center.
The sample is the largest, most complete LF every obtained for a
globular cluster. It is free from the problems of photometric blends
which plague ground-based LFs. In the crucial \msto/SGB region our
sample is more than three times larger than previous LFs.

Following a traditional LF function analysis we find no
surprises---canonical stellar models seem to fit the data well. In
particular, 

\begin{itemize}

\item The location of the RGB LF-bump agrees well new theoretical
models incorporating the OPAL opacities and enhancements of the
$\alpha$-elements. There is no indication of significant
``undershooting'' by the convective envelope. The best overall fit is
achieved with $\rm age=12$, $(m-M)_V = 15.03$ and global metallicity
$[M/H]\sim -1.2$ (equivalent to $\feh = -1.34$ and $\ofe = 0.2$).  On
the other hand, if we consider that the physics determining the bump
location might have more uncertainty than that determing  SGB
evolution, we might tolerate a worse fit for the bump to achieve a
better fit for the SGB. In that case we find $\rm
age=14$, $(m-M)_V = 14.80$ and $\metals \sim -1.2$

\item The slope of the integrated LF changes across the LF bump as
predicted by the models. This implies that the H-profile (and thus
fuel consumed in the earlier stages) is given reasonably accurately by
standard models.

\item The LF is well fit in the SGB region. There is no indication of
a problem with the canonical models as had been earlier found for low
metallicity clusters.

\item Below the TO we find evidence for a ``flat'' IMF ($x \approx 0$)
and evidence for mass segregation which is well fit by King models.

\end{itemize}

The two best observed GGC LFs, that of M3 presented here and M5 by SBSH,
seem to be a vindication of the quality of theoretical models. In
contrast, the anomalous LF of M30 first found by Bolte (1994) has
inconveniently persisted in later studies
(\cite{bergbuschm30,guham30,sandqm30}). Why should the theory work so well
in some cases and fail in others? One can explore possible deficiencies
in the models as in VandenBerg \etal\ (1998). However, we feel that the
ultimate resolution of the problem lies in the aquisition of more high quality
LFs to determine how common anomalous LFs are and what cluster parameters
they correlate with.

\acknowledgments

We thank Don VandenBerg for providing information prior to publication
and for very helpful email discussions. RTR \& BD are supported in part by NASA
Long Term Space Astrophysics Grant NAG 5-6403 and STScI/NASA Grants
GO-5969, GO-6804. This research was partially supported by the 
{\it Agenzia Spaziale Italiana} (ASI) and by the MURST 
as part of the project {\it Stellar Evolution}.
FRF acknowledges the {\it ESO Visiting Program} for its hospitality. 


\newpage
\section{FIGURE CAPTIONS}

\figcaption{ The $V$,~$V-I$ CMD for the HST global sample. The sample
contains more than 37,000 stars.
\label{hstcmd}}

\figcaption{ 
Map of the WFPC2 field of view
The circle (at $r \sim 50\arcsec$) delineates the INNER and OUTER
regions.
\label{WFPC2fld}}

\figcaption{
The smoothed interpolating 
curves of completeness as a function of $V$ magnitude.
\label{complt}}

\figcaption{
The zoomed LF with
and without completeness corrections. The Main-Sequence-Turn-Off point is labeled.
\label{hstlf}}

\figcaption{
CMDs of stars selected for the LF in each of the three adopted samples:
PHOTO, CCD, HST, respectively.
\label{3samplecmds}}

\figcaption{ Differential  Luminosity Function of the
global sample (PHOTO+CCD+HST). The names we use to refer to various
parts of the LF are indicated.
\label{lftot}}

\figcaption{
 Differential luminosity function of the global sample
compared with theoretical models from SCL97 (panel a)
and V97 (panel b). The chosen values for distance modulus, age and
chemical composition are indicated.
\label{cs-van}}

\figcaption{
 Integrated luminosity function of the global sample
compared with theoretical models by SCL97.
\label{lfint}}

\figcaption{The differential LF of the global sample
 is compared with SCL97 theoretical models varying metallicity, distance modulus
and age.
\label{scl}}

\figcaption{
The LF-bump region of the integrated  (panel a) and differential (panel b) LF:
The position of the LF-bump is indicated. The solid line is the
observed LF. The dashed line is based on SCL97 models. The short
dashed lines in (panel a) are straight line fits to the LF above and
below the LF-bump and are drawn to aid the eye in seeing the change in slope.
\label{bump}}

\figcaption{
The dependence of the SGB LF 	on WIMPs, He diffusion, $[\alpha/Fe]$.
\label{sgblf}}

\figcaption{A comparison of the LFs in three separate
radial regions: $r < 20\arcsec$; $50\arcsec < r < 100\arcsec$;
$210\arcsec < r < 360\arcsec$.
\label{lf3radii}}

\figcaption{
Comparison between the X value (computed as in Bolte 1989)
and the dynamical models from Pryor \etal\ 1986. 
\label{pryormod0}}

\newpage
\begin{deluxetable}{lcclccclc}
\tablewidth{\textwidth}
\label{ridgeline}
\tablecaption{Mean ridge line for the HST sample.}
\tablehead{
 \colhead{$V$}           & \colhead{$V-I$} &\colhead{}&
\colhead{$V$} & \colhead{$V-I$} &\colhead{} & \colhead{} &
\colhead{$V$} & \colhead{$V-I$}
}
\startdata
{\bf RGB} &  &   &  &   &    &    & {\bf HB}  & \nl
   12.44 &    1.589 &   &   18.16 &    0.803 &  &   &  15.57 &    0.775\nl
   12.68 &    1.473 &   &   18.26 &    0.787 &  &   &  15.62 &    0.731\nl
   12.90 &    1.403 &   &   18.33 &    0.764 &  &   &  15.63 &    0.684\nl
   13.10 &    1.347 &   &   18.37 &    0.746 &  &   &  15.64 &    0.631\nl
   13.30 &    1.294 &   &   18.42 &    0.724 &  &   &  15.64 &    0.585\nl
   13.55 &    1.241  &   &  18.47 &    0.696 &  &   &  15.62 &    0.272\nl
   13.79 &    1.193 &   &   18.52 &    0.668 &  &   &  15.63 &    0.219\nl
   14.05 &    1.150 &   &   18.57 &    0.647 &  &   &  15.64 &    0.184\nl
   14.28 &    1.115  &   &  18.62 &    0.629 &  &   &  15.66 &    0.146\nl
   14.56 &    1.077 &   &   18.70 &    0.616 &  &   &  15.71 &    0.105\nl
   14.82 &    1.045  &  &   18.78 &    0.608 &  &   &  15.76 &    0.079\nl
   15.09 &    1.017  &   &  18.86 &    0.600 &  &   &  15.83 &    0.043\nl
   15.36 &    0.989 &   &   18.98 &    0.593 &  &   &  15.91 &    0.008\nl
   15.64 &    0.964 &  &    19.10 &    0.592 &  &   &  15.99 &   -0.021\nl
   15.89 &    0.939 &   &   19.18 &    0.593 &  &   &  16.10 &   -0.050\nl
   16.21 &    0.911 &   &   19.26 &    0.595 &  &   &  16.21 &   -0.069\nl
   16.52 &    0.890  &   &  19.38 &    0.597 &  &   &  16.37 &   -0.080\nl
   16.70 &    0.879  &   &  19.78 &    0.624 &  &   &  16.54 &   -0.088\nl
   16.94 &    0.867  &   &  20.10 &    0.654 &  &   &  16.65 &   -0.091\nl
   17.20 &    0.856  &   &  20.30 &    0.672 &  &   &  &  \nl
   17.36 &    0.849  &   &  20.50 &    0.693 &  &    & &  \nl
   17.63 &    0.836 &   &   20.98 &    0.756 &  &    & &  \nl
   18.05 &    0.815  &   &  21.34 &    0.805 &  &   &  &  \nl
\enddata
\end{deluxetable}
\begin{deluxetable}{lccclcc}
\tablewidth{\textwidth}
\label{globallf}
\tablecaption{Luminosity function of the global sample (PHOTO+CCD+HST).}
\tablehead{
 \colhead{$V$}           & \colhead{$Log(N_i)$}      &
\colhead{$\sigma N_i$} &  \colhead{} & \colhead{$V$}&  
\colhead{$Log(N_i)$}&\colhead{$\sigma N_i$}
}
\startdata
    12.70 &     0.70 &     0.45 &  & 17.50 &     2.05 &     1.10\nl
    12.90 &     0.30 &     0.26 &  & 17.70 &     2.22 &     1.20\nl
    13.10 &     0.78 &     0.45 &  & 17.90 &     2.26 &     1.23\nl
    13.30 &     0.85 &     0.41 &  & 18.10 &     2.46 &     1.35\nl
    13.50 &     0.95 &     0.52 &  & 18.30 &     2.72 &     1.52\nl
    13.70 &     0.85 &     0.50 &  & 18.50 &     2.97 &     1.68\nl
    13.90 &     0.90 &     0.50 &  & 18.70 &     3.06 &     1.75\nl
    14.10 &     0.90 &     0.47 &  & 18.90 &     3.18 &     1.83\nl
    14.30 &     1.11 &     0.61 &  & 19.10 &     3.25 &     1.88\nl
    14.50 &     1.08 &     0.55 &  & 19.30 &     3.32 &     1.94\nl
    14.70 &     1.32 &     0.67 &  & 19.50 &     3.37 &     1.98\nl
    14.90 &     1.15 &     0.58 &  & 19.70 &     3.42 &     2.01\nl
    15.10 &     1.34 &     0.77 &  & 19.90 &     3.45 &     2.04\nl
    15.30 &     1.38 &     0.72 &  & 20.10 &     3.49 &     2.08\nl
    15.50 &     1.76 &     0.92 &  & 20.30 &     3.53 &     2.12\nl
    15.70 &     1.60 &     0.87 &  & 20.50 &     3.56 &     2.14\nl
    15.90 &     1.51 &     0.75 &  & 20.70 &     3.56 &     2.16\nl
    16.10 &     1.65 &     0.83 &  & 20.90 &     3.59 &     2.19\nl
    16.30 &     1.70 &     0.93 &  & 21.10 &     3.58 &     2.21\nl
    16.50 &     1.79 &     0.96 &  & 21.30 &     3.58 &     2.24\nl
    16.70 &     1.85 &     1.00 &  & 21.50 &     3.55 &     2.27\nl
    16.90 &     1.95 &     1.06 &  & 21.70 &     3.57 &     2.35\nl
    17.10 &     2.04 &     1.09 &  & 21.90 &     3.66 &     2.48\nl
    17.30 &     2.06 &     1.11 &  & &  & \
\enddata
\end{deluxetable}
\begin{deluxetable}{lccc}
\tablewidth{\textwidth}
\label{chi2}
\tablecaption{Chi-square results between observations and SCL models}
\tablehead{
 \colhead{}           & \colhead{$13<V<17$}      &
\colhead{$17<V<20.5$} & \colhead{$13<V<20.5$}
}
\startdata
$\rm [M/H]=-1.2 $& & &\nl
$(m-M)_V=15.03 $& & &\nl
$\rm age=12\,Gyr$& 0.101 & 0.014 & 0.114\nl
 & & & \nl
$\rm [M/H]=-1.2 $& & &\nl
$(m-M)_V=14.80 $& & &\nl
$\rm age=14\,Gyr$& 0.194 & 0.008 & 0.201\nl
 & & & \nl
$\rm  [M/H]=-0.82 $& & &\nl
$(m-M)_V=14.65 $& & &\nl
$\rm age=14\,Gyr$& 0.103 & 0.026 & 0.130\nl
 & & & \nl
$\rm [M/H]=-1.3 $& & &\nl
$(m-M)_V=15.20 $& & &\nl
$\rm age=10\,Gyr$& 0.112 & 0.018 & 0.131\nl
 & & & \nl

\enddata
\end{deluxetable}

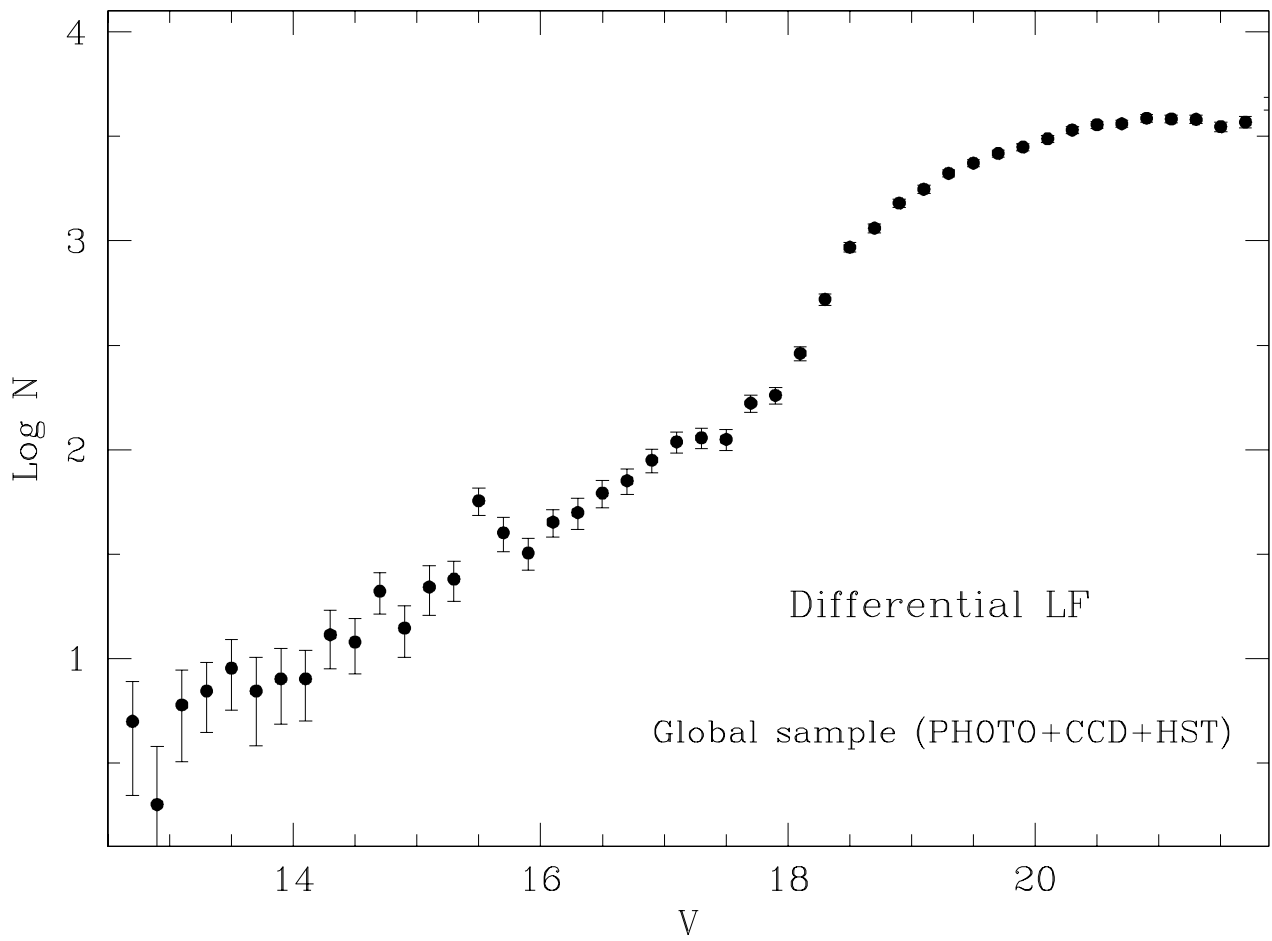
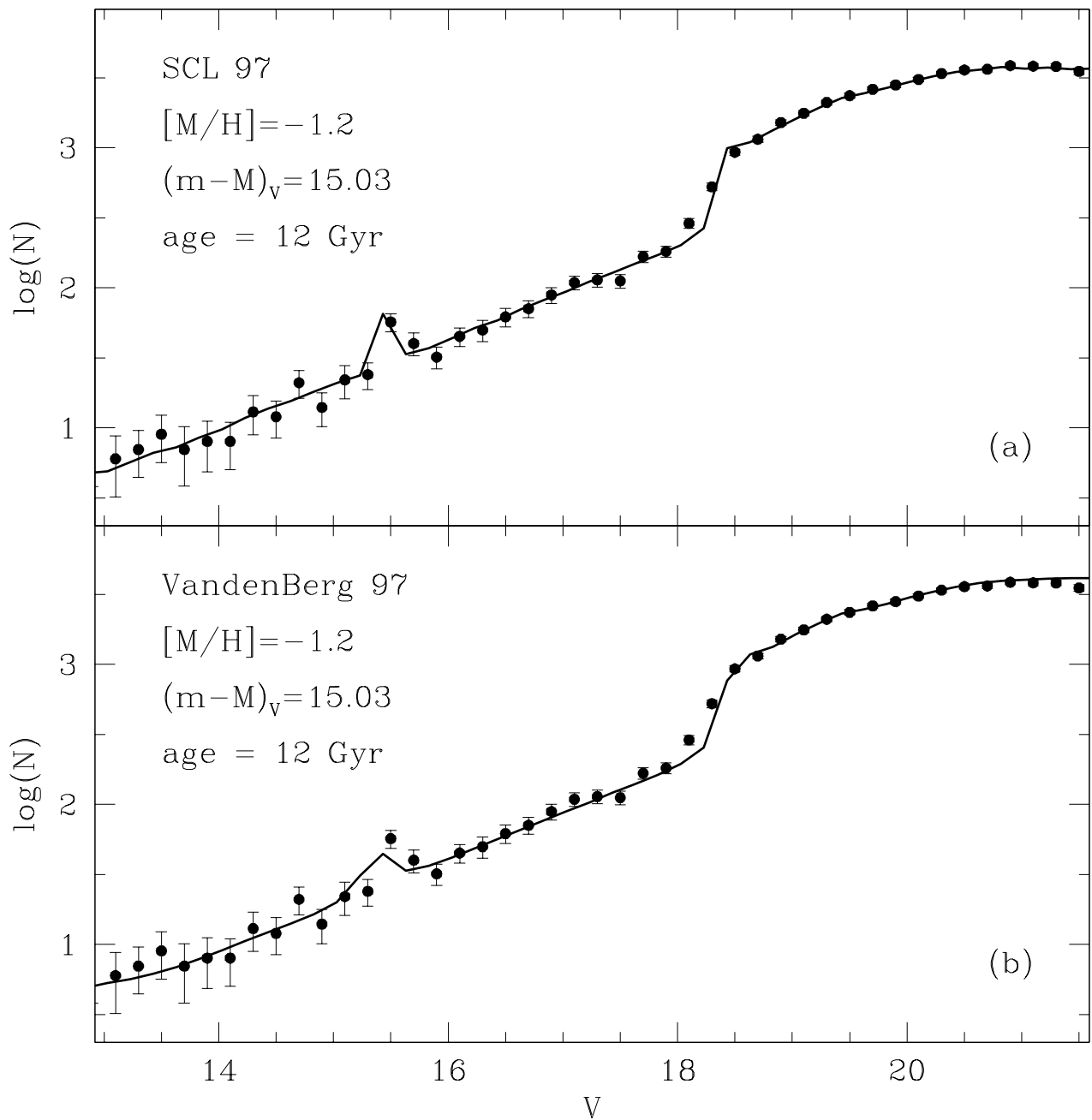
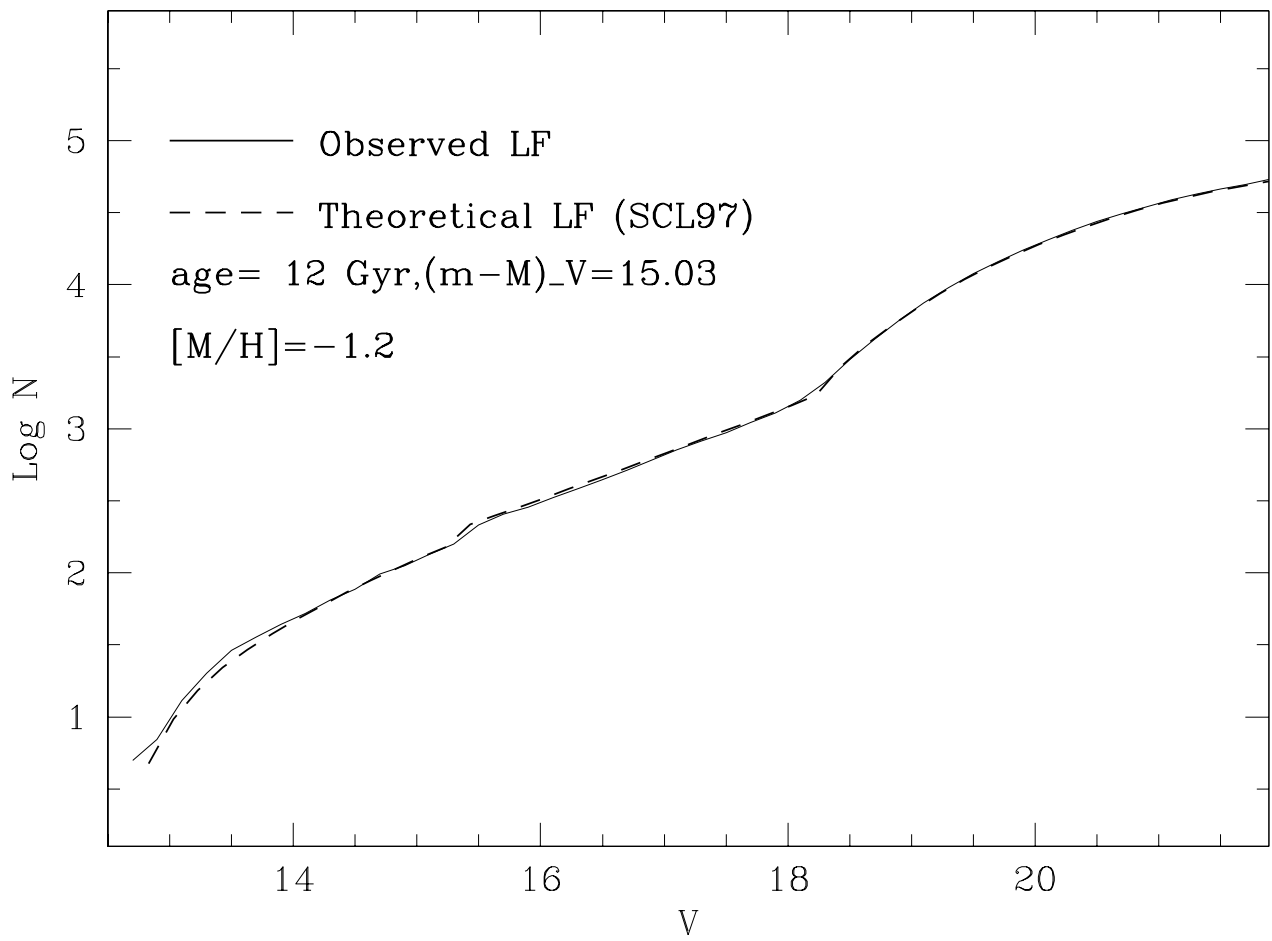
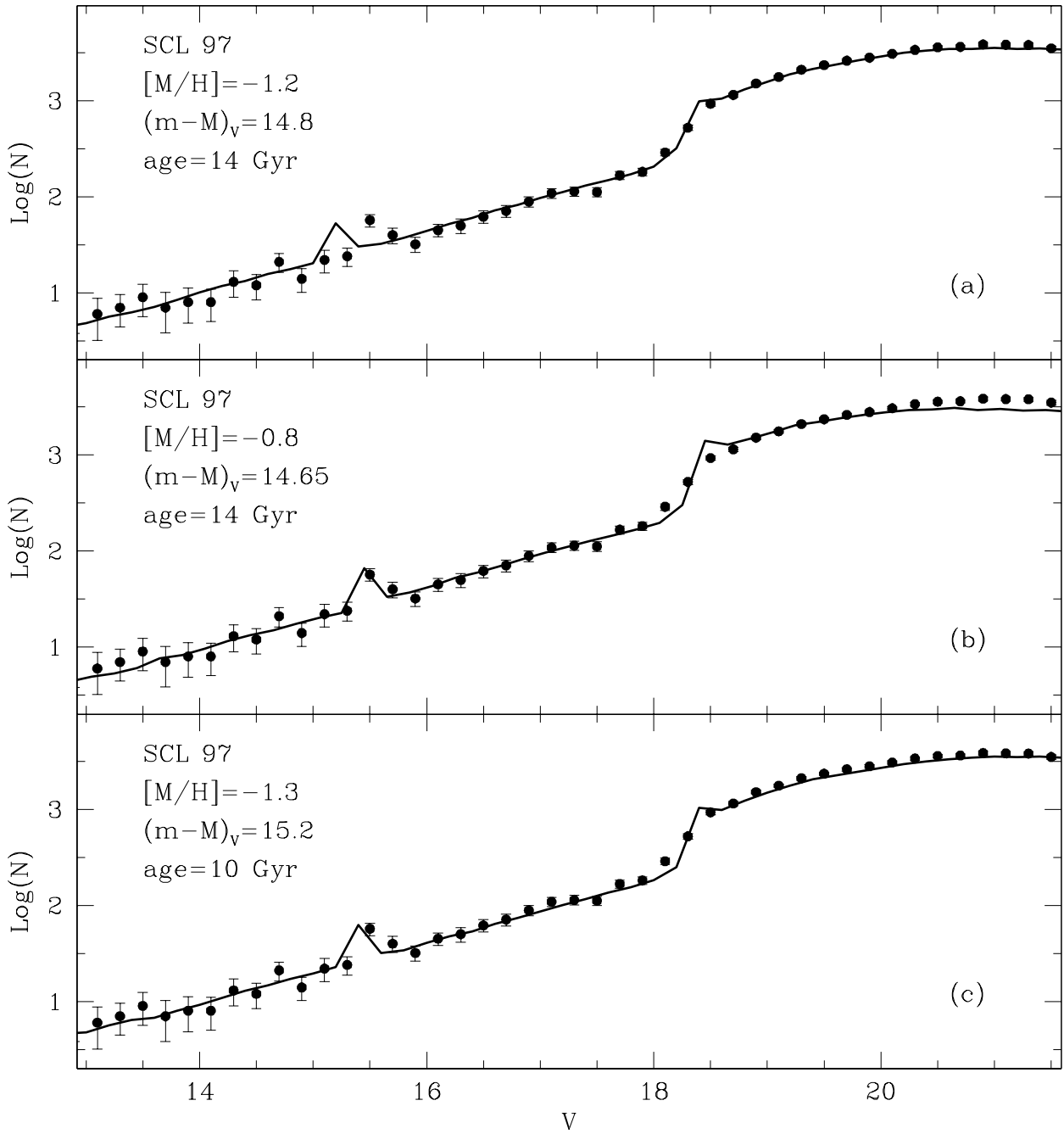
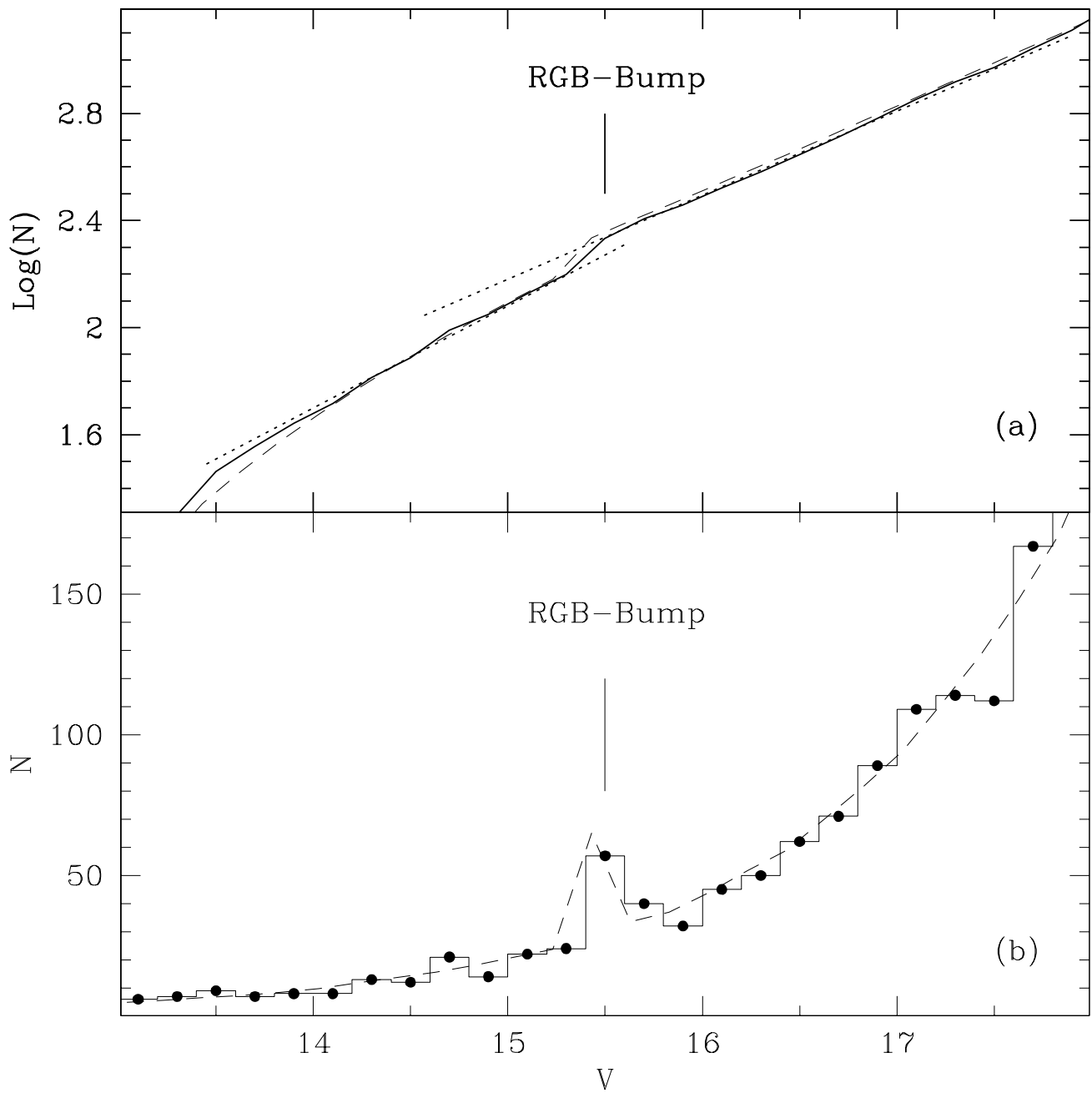
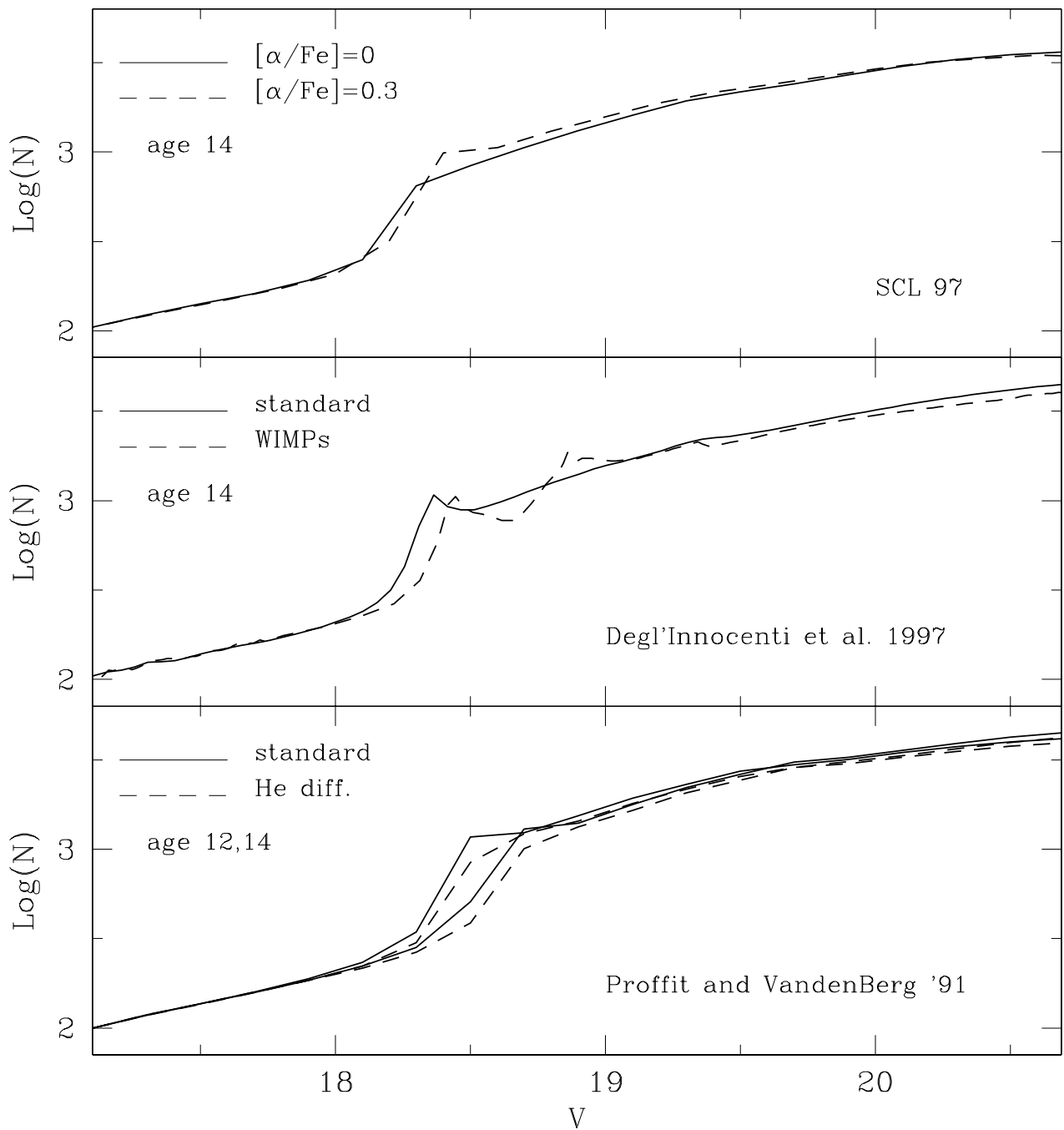
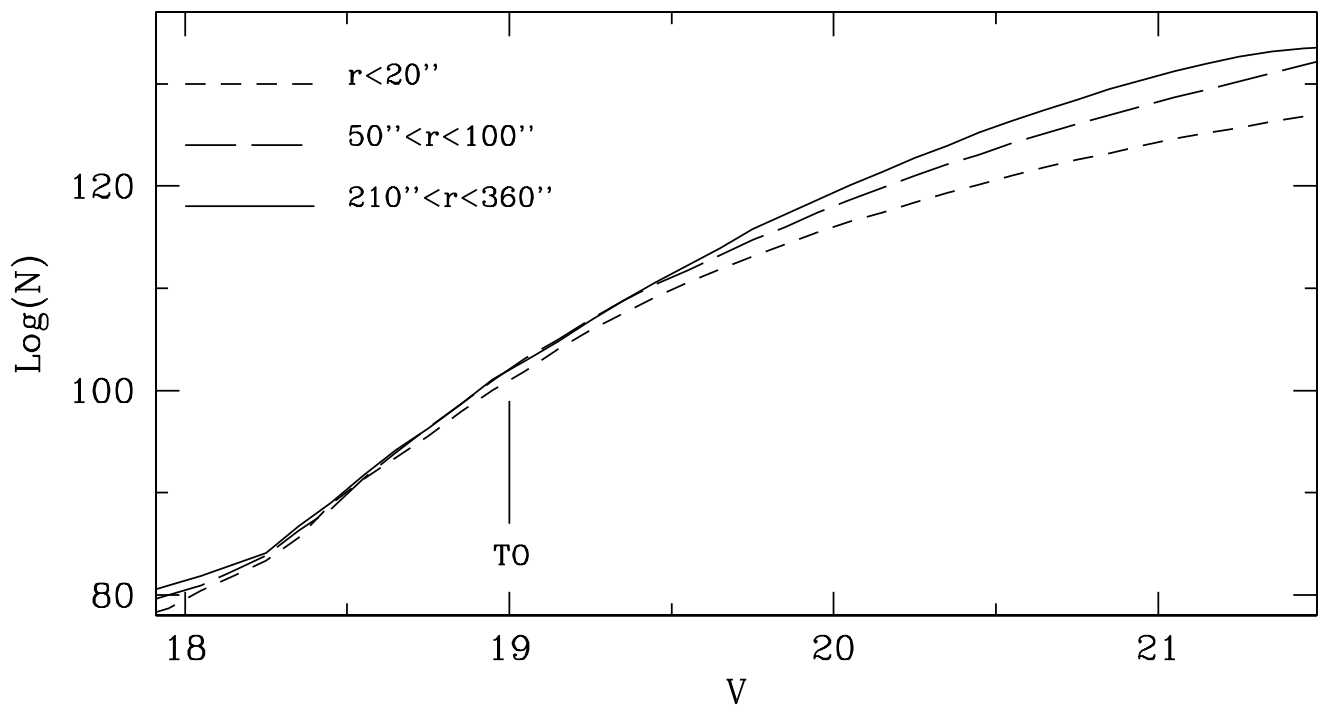
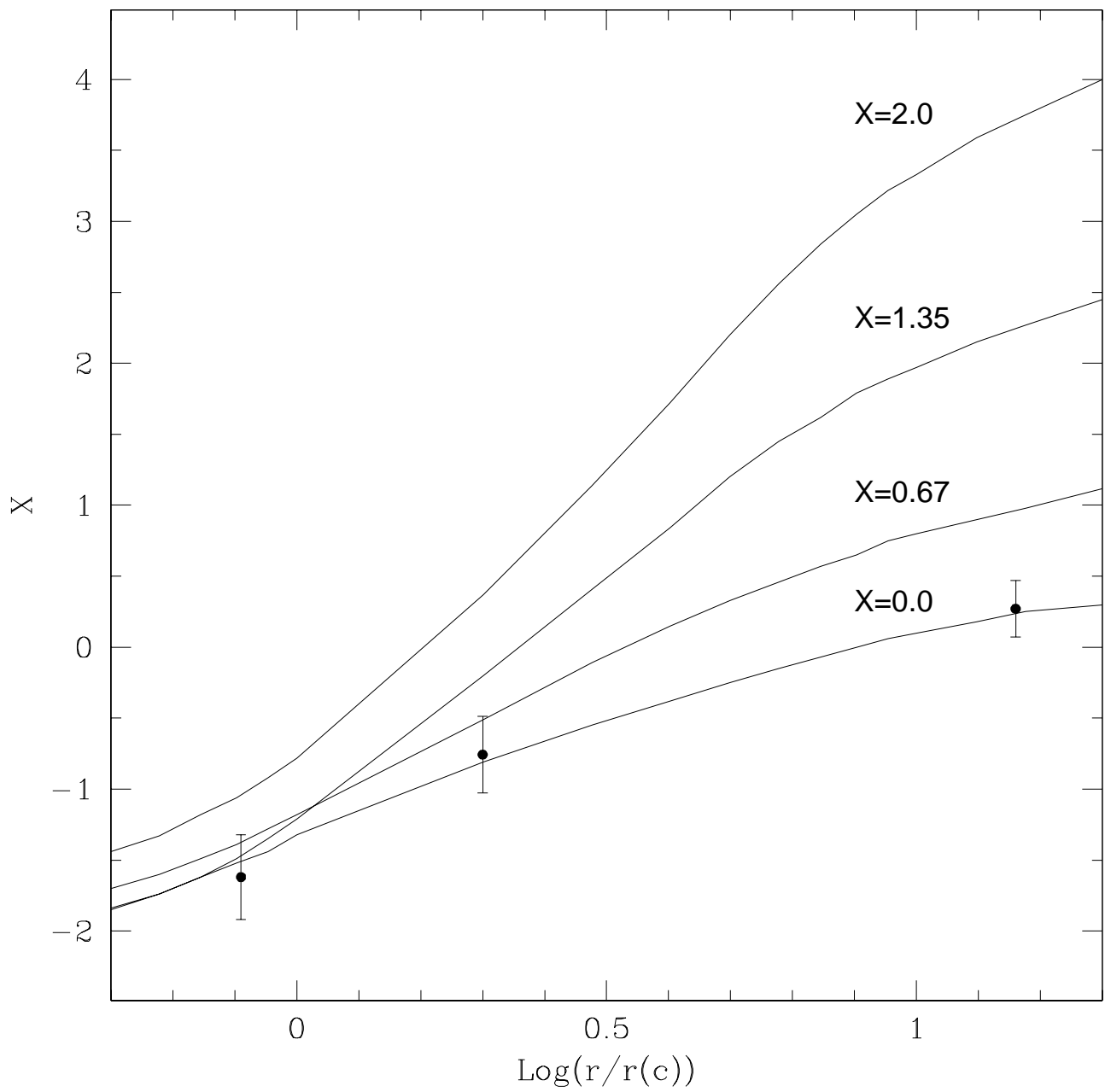
\end{document}